\documentclass[letter]{aa}
\usepackage{natbib}
\usepackage{graphicx}
\usepackage{txfonts}

\begin{document}

\title{The Amati relation in the ``fireshell'' model}

\author{
R. Guida\inst{1,2}
\and 
M.G. Bernardini\inst{1,2}
\and
C.L. Bianco\inst{1,2}
\and
L. Caito\inst{1,2}
\and
M.G. Dainotti\inst{1,2}
\and
R. Ruffini\inst{1,2,3}
}

\institute{
ICRANet and ICRA, Piazzale della Repubblica 10, I-65122 Pescara, Italy.
\and
Dipartimento di Fisica, Universit\`a di Roma ``La Sapienza'', Piazzale Aldo Moro 5, I-00185 Roma, Italy. E-mails: roberto.guida@icra.it, maria.bernardini@icra.it, bianco@icra.it, letizia.caito@icra.it, dainotti@icra.it, ruffini@icra.it.
\and
ICRANet, Universit\'e de Nice Sophia Antipolis, Grand Ch\^ateau, BP 2135, 28, avenue de Valrose, 06103 NICE CEDEX 2, France.
}

\titlerunning{The Amati relation in the ``fireshell'' model}

\authorrunning{Guida et al.}

\date{}

\abstract{
The cosmological origin of Gamma-Ray Bursts (GRBs) has been firmly established, with redshifts up to $z = 6.29$. GRBs are possible candidates to be used as ``distance indicators'' to test cosmological models in a redshift range hardly achievable by other cosmological probes. Asserting the validity of the empirical relations among GRB observables is now crucial for their calibration.
}{
Motivated by the relation proposed by Amati and collaborators, we look within the ``fireshell'' model for a relation between the peak energy $E_p$ of the $\nu F_\nu$ total time-integrated spectrum of the afterglow and the total energy of the afterglow $E_{aft}$, which in our model encompasses and extends the prompt emission.
}{
The fit within the fireshell model, as for the ``canonical'' GRB050315, uses the complete arrival time coverage given by the Swift satellite. It is performed simultaneously, self-consistently and recursively in the four BAT energy bands ($15$--$25$ keV, $25$--$50$ keV, $50$--$100$ keV and $100$-$150$ keV) as well as in the XRT one ($0.2$--$10$ keV). It uniquely determines the two free parameters characterizing the GRB source, the total energy $E^{e^\pm}_{tot}$ of the $e^{\pm}$ plasma and its baryon loading $B$, as well as the effective CircumBurst medium (CBM) distribution. We can then build two sets of ``gedanken'' GRBs varying the total energy of the electron-positron plasma $E^{e^\pm}_{tot}$ and keeping the same baryon loading $B$ of GRB050315. The first set assumes for the effective CBM density the one obtained in the fit of GRB050315. The second set assumes instead a constant CBM density equal to the average value of the GRB050315 prompt phase.
}{
For the first set of ``gedanken'' GRBs we find a relation $E_p\propto (E_{aft})^a $, with $a = 0.45 \pm 0.01$, whose slope strictly agrees with the Amati one. Such a relation, in the limit $B \to 10^{-2}$, coincides with the Amati one. Instead, in the second set of ``gedanken'' GRBs no correlation is found.
}{
Our analysis excludes the Proper-GRB (P-GRB) from the prompt emission, extends all the way to the latest afterglow phases and is independent on the assumed cosmological model, since all ``gedanken'' GRBs are at the same redshift. The Amati relation, on the other hand, includes also the P-GRB, focuses on the prompt emission only, and is therefore influenced by the instrumental threshold which fixes the end of the prompt emission, and depends on the assumed cosmology. This may well explain the intrinsic scatter observed in the Amati relation.}


\maketitle

\section{Introduction}\label{intro}

The detection of Gamma-Ray Bursts (GRBs) up to very high redshifts \citep[up to $z = 6.29$, see][]{2005A&A...443L...1T}, their high observed rate of one every few days, and the progress in the theoretical understanding of these sources make them useful as cosmological tools, complementary to Supernovae Ia which are observed only up to $z = 1.7$ \citep{2001ARA&A..39...67L,2001ApJ...560...49R}. One of the hottest topics on GRBs is constituted by the possible existence of empirical relations between GRB observables \citep{2002A&A...390...81A,2004ApJ...616..331G,2004ApJ...609..935Y,2005ApJ...633..611L,2006MNRAS.370..185F,2008arXiv0805.0377A}, which may lead, if confirmed, to use GRBs as tracers of models of universe. The first empirical relation, discovered analyzing the \emph{BeppoSAX} so-called ``long'' bursts with known redshift, was the ``Amati relation'' \citep{2002A&A...390...81A}. It was found that the isotropic-equivalent radiated energy of the prompt emission $E_{iso}$ is correlated with the cosmological rest-frame $\nu F_{\nu}$ spectrum peak energy $E_{p,i}$: $E_{p,i}\propto (E_{iso})^{a}$, with $a = 0.52 \pm 0.06$ \citep{2002A&A...390...81A}. The existence of the Amati relation has been confirmed by studying a sample of GRBs discovered by Swift, with $a = 0.49^{+0.06}_{-0.05}$ \citep{2006ApJ...636L..73S,2006MNRAS.372..233A}.

Swift has given for the first time the possibility to obtain high quality data in selected energy bands from the GRB trigger time all the way to the latest afterglow phases \citep{2004ApJ...611.1005G}. This has given the opportunity to apply our theoretical ``fireshell'' model obtaining detailed values for its two free parameters, the total energy $E^{e^\pm}_{tot}$ and the baryon loading $B$ of the fireshell, as well as the values of the effective density and filamentary structure of the CircumBurst Medium (CBM). This allowed to compute multi-band light curves and spectra, both instantaneous and time-integrated, compared with selected GRB sources, like e.g. GRB050315.

In the ``fireshell'' model $E^{e^\pm}_{tot}$ comprises two different components: the Proper GRB (P-GRB), with energy $E_{P-GRB}$, emitted at the moment when the $e^+e^-$-driven accelerating baryonic matter reaches transparency, and the following afterglow phase, with energy $E_{aft}$, with the decelerating baryons interacting with the CBM \citep{2001ApJ...555L.113R}. These two phases are clearly distinguishable by their relative intensity and temporal separation in arrival time. We have:
\begin{equation}
E^{e^\pm}_{tot} = E_{P-GRB} + E_{aft}\, .
\label{Etot}
\end{equation}
What is usually called the ``prompt emission'' corresponds within the fireshell model to the P-GRB together with the peak of the afterglow \citep[see below, e.g.][and references therein]{2001ApJ...555L.113R,2006ApJ...645L.109R,2007AIPC..910...55R,2007A&A...471L..29D,2007A&A...474L..13B,2008AIPC..966...16C,2008AIPC..966...12B}.

Among the crucial issues raised by the Amati relation there are its theoretical explanation and its possible dependence on the assumed cosmological model. We have examined a set of ``gedanken'' GRBs, all at the same cosmological redshift of GRB050315. Such a set assumes the same fireshell baryon loading and effective CBM distribution as GRB050315 and each ``gedanken'' GRB differs from the others uniquely by the value of its total energy $E^{e^\pm}_{tot}$. We have then considered a second set of ``gedanken'' GRBs, differing from the previous one by assuming a constant effective CBM density instead of the one inferred for GRB050315. In both these sets we looked for a relation between the isotropic-equivalent radiated energy of the \emph{entire} afterglow $E_{aft}$, and the corresponding time-integrated $\nu F_{\nu}$ spectrum peak energy $E_p$:
\begin{equation}
E_p\propto (E_{aft})^a\, .
\label{al}
\end{equation}

In section \ref{model} we recall the main features of the ``fireshell'' model. In section \ref{fit} we recall the main features of the GRB050315 fitting procedure. In section \ref{correlation} we then present the derivation of the $E_p$ -- $E_{aft}$ relation for two sets of ``gedanken'' GRBs. The results of this analysis and their discussion are shown in section \ref{results}. In section \ref{conclusions} we present the conclusions, strongly confirming the validity of the Amati relation.

\section{The ``fireshell'' model and the canonical GRB scenario}\label{model}

Our ``fireshell'' model assumes that all GRBs originate from the gravitational collapse to a black hole \citep{2001ApJ...555L.113R,2007AIPC..910...55R}. The $e^\pm$ plasma created in the process of the black hole formation expands as a spherically symmetric ``fireshell'' with a constant width of the order of $\sim 10^8$ cm in the laboratory frame, i.e. the frame in which the black hole is at rest. We have only two free parameters characterizing the source, namely the total energy $E^{e^\pm}_{tot}$ of the $e^\pm$ plasma and its baryon loading $B\equiv M_Bc^2/E^{e^\pm}_{tot}$, where $M_B$ is the total baryons' mass \citep{2000A&A...359..855R}. They fully determine the optically thick acceleration phase of the fireshell, which lasts until the transparency condition is reached and the Proper-GRB (P-GRB) is emitted \citep{2001ApJ...555L.113R}. We recall that in the current literature the P-GRB is sometimes considered a ``precursor'' of the main GRB event.

After this optically thick acceleration phase, the optically thin deceleration phase starts, with the afterglow emission due to the collision between the remaining accelerated baryonic matter and the CBM. This emission clearly depends on the parameters describing the effective CBM distribution: its density $n_{cbm}$ and the ratio ${\cal R}\equiv A_{eff}/A_{vis}$ between the effective emitting area of the fireshell $A_{eff}$ and its total visible area $A_{vis}$ \citep{2002ApJ...581L..19R,2004IJMPD..13..843R,2005IJMPD..14...97R}. The radiation emitted during the afterglow is assumed to have a thermal spectrum in the co-moving frame of the fireshell. Due to the temporal evolution of the fireshell temperature and to the Doppler effect implied by its ultra-relativistic expansion, the observed afterglow spectra are non-thermal because they are convolutions of thousands of thermal spectra with different temperatures over the corresponding EQuiTemporal Surfaces \citep[EQTSs, the surfaces of equal arrival time of the photons at the detector, see][]{2004ApJ...605L...1B,2005ApJ...620L..23B} and over the observation time \citep{2004IJMPD..13..843R,2005ApJ...634L..29B}.

We indeed define within our model the ``canonical GRB'' light curve as made by two sharply physical different components: the P-GRB and the afterglow \citep{2001ApJ...555L.113R,2007AIPC..910...55R,2008AIPC..966...12B}. The former has the imprint of the black hole formation, an harder spectrum and no spectral lag \citep{2001A&A...368..377B,2005IJMPD..14..131R}. The latter presents a clear hard-to-soft spectral evolution in time, and it consists of three well defined different regimes: a rising branch, a peak, and a decaying tail \citep{2001ApJ...555L.113R}. The ratio between the total time-integrated luminosities (i.e. the total energies) of the P-GRB and the afterglow as well as their arrival time separation are ruled by $E^{e^\pm}_{tot}$ and $B$ \citep{2000A&A...359..855R,2001ApJ...555L.113R}. When $B \lesssim 10^{-5}$, the P-GRB is the leading contribution to the emission and the afterglow is negligible: we have a ``genuine'' short GRB \citep{2001ApJ...555L.113R,2007A&A...474L..13B,2008AIPC..966...12B}. When $10^{-4} \lesssim B \lesssim 10^{-2}$, instead, the afterglow contribution is generally predominant \citep[for the existence of the upper limit $B \lesssim 10^{-2}$ see][]{2000A&A...359..855R}. Still, this last case presents two distinct possibilities: the afterglow peak luminosity can be either \emph{larger} or \emph{smaller} than the P-GRB one. This last case, i.e. an afterglow with total time-integrated luminosity larger than the P-GRB one but with a smaller peak luminosity, is indeed explainable in terms of a peculiarly small average value of the CBM density ($n_{cbm} \sim 10^{-3}$ particles/cm$^3$), compatible with a galactic halo environment. Such a small average CBM density ``deflates'' the afterglow peak luminosity which, therefore, has to last longer since its total energy is fixed by the value of $B$. This is the class of the ``fake'' short GRBs, which present an initial spikelike emission followed by a prolonged soft bump \citep[see][and references therein]{2007A&A...474L..13B,2008AIPC..966....7B,2008AIPC..966...12B} and includes the sources analyzed by \citet{2006ApJ...643..266N}.

What is usually called the ``prompt emission'' neglects all the above analysis: it comprises in fact the P-GRB and the first two regimes of the afterglow, namely the rising branch and the peak. What is usually called the ``afterglow'' is just the third regime, namely its decaying tail \citep[see e.g.][]{2005ApJ...634L..29B,2006ApJ...645L.109R,2007AIPC..910...55R,2007A&A...471L..29D,2007A&A...474L..13B,2008AIPC..966...12B}. Within our approach there is no way to find a separation between the end of the ``prompt emission'' and the beginning of the decaying tail of the afterglow. For these reasons, in the following, we analyze a relation between the isotropic-equivalent radiated energy of the \emph{entire} afterglow $E_{aft}$, and the corresponding time-integrated $\nu F_{\nu}$ spectrum peak energy $E_p$.

\section{GRB050315 best fit}\label{fit}

In \citet{2006ApJ...645L.109R} we analyzed GRB050315. Thanks to the high quality gapless data \citep{2006ApJ...638..920V} provided by the Swift satellite \citep{2004ApJ...611.1005G}, the fit of the observed GRB050315 light curves had to be performed simultaneously, self-consistently and recursively in the four BAT energy bands ($15$--$25$ keV, $25$--$50$ keV, $50$--$100$ keV and $100$-$150$ keV) as well as in the XRT one ($0.2$--$10$ keV). This fitting procedure fixed the values of $E^{e^\pm}_{tot}$, of $B$ and of the effective CBM distribution. It is important here to emphasize that such values are very tightly constrained by the observational data. In particular, each single value of the CBM density mask affects in a nonlinear way all the subsequent evolution of the light curves, though both the dynamical equations and the EQTSs, thus making any ``piecewise'' analysis of the light curves impossible.

\begin{figure}
\includegraphics[width=\hsize,clip]{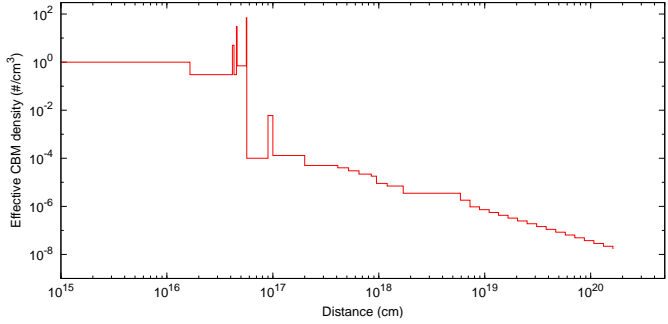}
\caption{The effective CBM number density inferred from the theoretical analysis of GRB050315. Details in \citet{2006ApJ...645L.109R}.}
\label{mask}
\end{figure}

For GRB050315 we obtained $E^{e^\pm}_{tot} = 1.46\times 10^{53}$ erg and $B = 4.55 \times 10^{-3}$ \citep{2006ApJ...645L.109R}. These two values determine the fireshell dynamics up to the transparency, and in particular they fix the ratio between $E_{P-GRB}$ and $E_{aft}$ and the temporal separation, measured in detector arrival time, between the P-GRB and the peak of the afterglow \citep{2001ApJ...555L.113R,2003AIPC..668...16R,2006ApJ...645L.109R}. The effective CBM density profile inferred from our theory for GRB050315 is shown in Fig. \ref{mask} \citep{2006ApJ...645L.109R}. Such values and such a profile, as stated above, have been obtained fitting the five BAT and XRT light curves of the entire GRB.

\section{The $E_p$ -- $E_{aft}$ relation}\label{correlation}

In our approach only the \emph{entire} afterglow emission is considered in establishing our $E_p$ -- $E_{aft}$ relation. From this assumption one derives, in a natural way, the fact that the Amati relation holds only for long GRBs, where the P-GRB is negligible, and not for short GRBs \citep{2007A&A...463..913A}.

Within our theoretical model we can compute the ``instantaneous'' spectrum of GRB050315 at each value of the detector arrival time during the entire afterglow emission. Such a spectrum sharply evolve in arrival time presenting a typical hard-to-soft behavior \citep{2006ApJ...645L.109R}. We have then computed the $\nu F_\nu$ time-integrated spectrum over the total duration of our afterglow phase, that is, from the end of the P-GRB up to when the fireshell reaches a Lorentz gamma factor close to unity. We can then define the energy $E_p$ as the energy of the peak of such $\nu F_\nu$ time-integrated spectrum and we look to its relation with the total energy $E_{aft}$ of the afterglow.

We construct, at a fixed cosmological redshift, therefore independently of the cosmological model, two sets of ``gedanken'' GRBs. The first set assumes the same fireshell baryon loading and effective CBM distribution as GRB050315 (see Fig. \ref{mask}) and each ``gedanken'' GRB differs from the others uniquely by the value of its total energy $E^{e^\pm}_{tot}$. The second set assumes a constant effective CBM density $\sim 1$ particle/cm$^3$ instead of the one inferred for GRB050315.

\begin{figure}
\includegraphics[width=\hsize,clip]{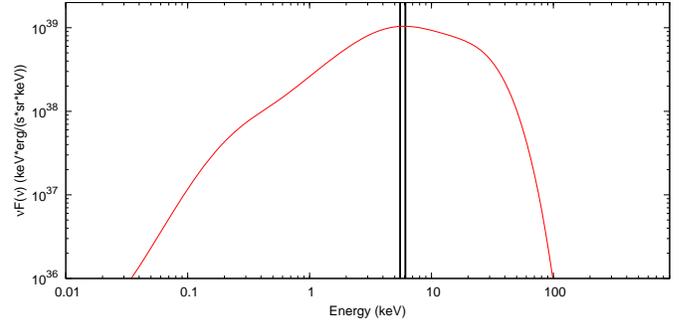}
\caption{The $\nu F_\nu$ time-integrated spectrum over the total duration of our afterglow phase for the ``gedanken'' GRB of the first set with total energy $E^{e^\pm}_{tot} = 3.40\times 10^{51}$. The two vertical lines constrain the $5\%$ error region around the peak. We determine $E_p = 5.82$ keV $\pm 5\%$.}
\label{maximum_spectrum}
\end{figure}

Within our model $E_{aft}$ is a fixed value determined by $E^{e^\pm}_{tot}$ and $B$, so clearly there are no errors associated to it. Instead, $E_p$ is evaluated from the numerically calculated spectrum, and its determination is therefore affected by the numerical resolution. Choosing a $5\%$ error on $E_p$, consistent with our numerical resolution, we checked, looking at each spectrum, that this value is reasonable. Fig. \ref{maximum_spectrum} shows the time integrated spectrum corresponding to $E^{e^\pm}_{tot} = 3.40\times 10^{51}$ erg with the error around $E_p$.

\section{Results and discussion}\label{results}

\begin{figure}
\includegraphics[width=\hsize,clip]{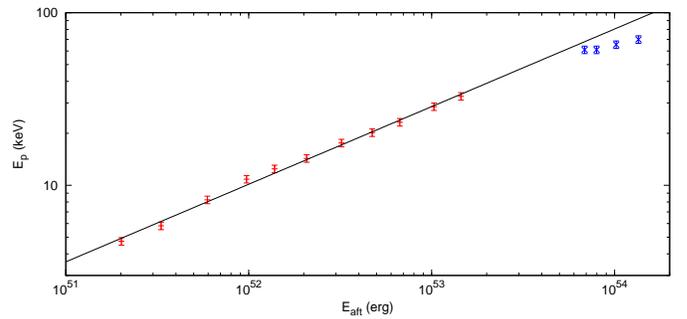}
\caption{The $E_p$ -- $E_{aft}$ relation: the results of the simulations of the first set of ``gedanken'' GRBs (red points) are well fitted by a power-law (black line) $E_p\propto (E_{aft})^a$ with $a = 0.45 \pm 0.01$. Such a power-law slope strictly agrees with the Amati relation. We also see that, extending the first set of ``gedanken'' GRBs to $E^{e^\pm}_{tot} > 10^{53}$ erg and restricting the $E_p$ determination to the high-energy spectral peak (blue points), the same relation is still fulfilled for $E^{e^\pm}_{tot} \sim 10^{54}$ erg. We notice a possible saturation for $E^{e^\pm}_{tot} > 10^{54}$ erg.}
\label{fig-correlation_extended}
\end{figure}

Fig. \ref{fig-correlation_extended} shows the $E_p$ -- $E_{aft}$ relation of the ``gedanken'' GRBs belonging to the first set (red points). It extends over two orders of magnitude in energy, from $10^{51}$ to $10^{53}$ erg, and is well fitted by a power-law $E_p\propto (E_{aft})^a$ with $a = 0.45 \pm 0.01$. We emphasize that such a power-law slope strictly agrees with the Amati relation, namely $E_{p,i}\propto (E_{iso})^{a}$, with $a = 0.49 ^{+0.06}_{-0.05}$ \citep{2006MNRAS.372..233A}. We recall that $E_p$ is the observed peak energy, namely it is not rescaled for the cosmological redshift, because all the ``gedanken'' GRBs of the set are at the same redshift of GRB050315, namely $z=1.949$ \citep{2006ApJ...638..920V}. The normalization is clearly different from the Amati one.

\begin{figure}
\includegraphics[width=\hsize,clip]{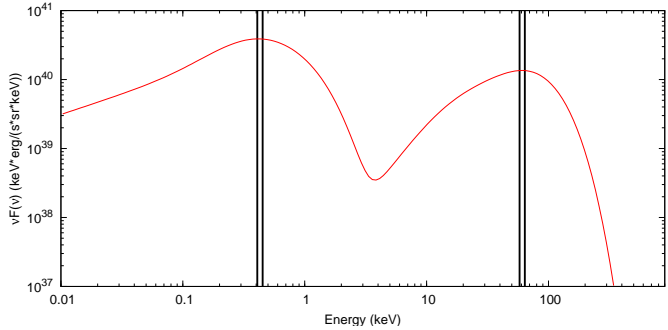}
\caption{The $\nu F_\nu$ time-integrated spectrum over the total duration of our afterglow phase for the ``gedanken'' GRB of the extended first set with total energy $E^{e^\pm}_{tot} = 6.95\times 10^{53}$ erg. The vertical lines constrain the $5\%$ error region around each peak.}
\label{spectrum_G_2_peaks}
\end{figure}

If we try to extend the first sample of ``gedanken'' GRBs below $10^{51}$ erg, the relevant CBM distribution would be for $r \lesssim 10^{16}$ cm, where no data are available from the GRB050315 observations. If we try to extend the first set of ``gedanken'' GRBs above $10^{53}$ erg, we notice that for $E^{e^\pm}_{tot} \gtrsim 10^{54}$ erg the small ``bump'' which can be noticed between $0.2$ and $1.0$ keV in the spectrum of Fig. \ref{maximum_spectrum} evolves into a low-energy second spectral peak which is even higher than the high-energy one (see Fig. \ref{spectrum_G_2_peaks}). We are currently investigating if this low-energy second peak is a real theoretically predicted spectral feature which may be observed in the future in highly energetic sources. There is also the other possibility that the low-energy and late times part of our GRB050315 fit is not enough constrained by the XRT observational data and that this effect is magnified by the $E^{e^\pm}_{tot}$ rescaling. 

The high-energy spectral peak is due to the emission at the peak of the afterglow, and therefore due to the so-called ``prompt emission''. The low-energy one is due to late times soft X-ray emission. Therefore, the high-energy spectral peak is the relevant one for the Amati relation. We find indeed that such an high-energy spectral peak still fulfills the $E_p$ -- $E_{aft}$ relation for $E^{e^\pm}_{tot} \sim 10^{54}$ erg, with a possible saturation for $E^{e^\pm}_{tot} > 10^{54}$ erg (see blue points in Fig. \ref{fig-correlation_extended}).

\begin{figure}
\includegraphics[width=\hsize,clip]{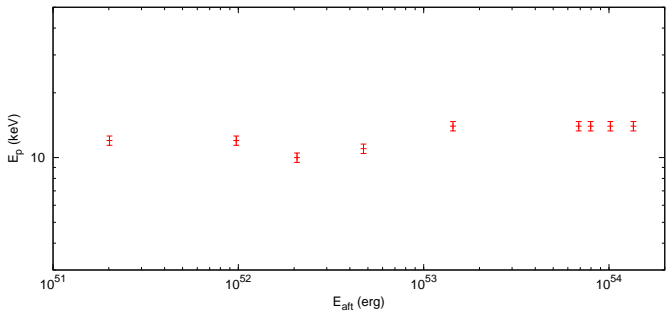}
\caption{The second set of ``gedanken'' GRBs. Clearly, in this case there in no relation between $E_p$ and $E_{aft}$.}
\label{correlation-no-mask}
\end{figure}

Fig. \ref{correlation-no-mask} clearly shows that in the second set of ``gedanken'' GRBs, built assuming a constant effective CBM density $\sim 1$ particle/cm$^3$, instead of the one specifically inferred for GRB050315, there in no relation between $E_p$ and $E_{aft}$.

\section{Conclusions}\label{conclusions}

The Swift high quality data, giving for the first time gapless and multiwavelength coverage from the GRB trigger all the way to the latest afterglow phases, have allowed to make a complete fit of the GRB050315 multiband light curves based on our fireshell model. We have so fixed the free parameters describing the source and we determined the instantaneous and time-integrated spectra during the entire afterglow.

Starting from this, we have examined two sets of ``gedanken'' GRBs, constructed at a fixed cosmological redshift. The first set assumes the same fireshell baryon loading and effective CBM distribution as GRB050315 and each ``gedanken'' GRB differs from the others uniquely by the value of its total energy $E^{e^\pm}_{tot}$. The second set assumes a constant effective CBM density $\sim 1$ particle/cm$^3$ instead of the one inferred for GRB050315.

Recalling that the ``canonical'' GRB light curve in the fireshell model is composed of two well separated components, the P-GRB and the entire afterglow, we have looked in both sets for a relation between the isotropic-equivalent radiated energy of the \emph{entire} afterglow $E_{aft}$ and the corresponding time-integrated $\nu F_{\nu}$ spectrum peak energy $E_p$: $E_p\propto (E_{aft})^a$. In doing so, we have assumed that the Amati relation is directly linked to the interaction between the accelerated baryons and the CBM. The P-GRBs, which originates from the fireshell transparency, in our approach do not fulfill the Amati relation. Consequently, also the short GRBs, which have a vanishing afterglow with respect to the P-GRB, should not fulfill the Amati relation. This last point is supported by the observational evidence \citep{2006MNRAS.372..233A}.

We notice that the first set of ``gedanken'' GRBs fulfills very well the $E_p\propto (E_{aft})^a$ relation with $a = 0.45 \pm 0.01$. This slope strongly agrees with the Amati relation. On the contrary, for the second set no relation between $E_p$ and $E_{aft}$ seems to hold. We conclude that the Amati relation originates from the detailed structure of the effective CBM.

Turning now to the analogies and differences between our $E_p$ -- $E_{aft}$ relation and the Amati one, our analysis excludes the Proper-GRB (P-GRB) from the prompt emission, extends all the way to the latest afterglow phases and is independent on the assumed cosmological model, since all ``gedanken'' GRBs are at the same redshift. The Amati relation, on the other hand, includes also the P-GRB, focuses on the prompt emission only, and is therefore influenced by the instrumental threshold which fixes the end of the prompt emission, and depends on the assumed cosmology. This may well explain the intrinsic scatter observed in the Amati relation \citep{2006MNRAS.372..233A}. Our theoretical work is a first unavoidable step to support the usage of the empirical Amati relation for measuring the cosmological parameters.

\acknowledgements

We thank the anonymous referee for his/her very constructive advices.


\begin{thebibliography}{33}
\expandafter\ifx\csname natexlab\endcsname\relax\def\natexlab#1{#1}\fi

\bibitem[{{Amati}(2006)}]{2006MNRAS.372..233A}
{Amati}, L. 2006, MNRAS, 372, 233

\bibitem[{{Amati} {et~al.}(2007){Amati}, {Della Valle}, {Frontera}, {Malesani},
  {Guidorzi}, {Montanari}, \& {Pian}}]{2007A&A...463..913A}
{Amati}, L., {Della Valle}, M., {Frontera}, F., {et~al.} 2007, A\&A, 463, 913

\bibitem[{{Amati} {et~al.}(2002){Amati}, {Frontera}, {Tavani}, {in't Zand},
  {Antonelli}, {Costa}, {Feroci}, {Guidorzi}, {Heise}, {Masetti}, {Montanari},
  {Nicastro}, {Palazzi}, {Pian}, {Piro}, \& {Soffitta}}]{2002A&A...390...81A}
{Amati}, L., {Frontera}, F., {Tavani}, M., {et~al.} 2002, A\&A, 390, 81

\bibitem[{{Amati} {et~al.}(2008){Amati}, {Guidorzi}, {Frontera}, {Della Valle},
  {Finelli}, {Landi}, \& {Montanari}}]{2008arXiv0805.0377A}
{Amati}, L., {Guidorzi}, C., {Frontera}, F., {et~al.} 2008, ArXiv:0805.0377

\bibitem[{{Bernardini} {et~al.}(2007){Bernardini}, {Bianco}, {Caito},
  {Dainotti}, {Guida}, \& {Ruffini}}]{2007A&A...474L..13B}
{Bernardini}, M.~G., {Bianco}, C.~L., {Caito}, L., {et~al.} 2007, A\&A, 474,
  L13

\bibitem[{{Bernardini} {et~al.}(2008){Bernardini}, {Bianco}, {Caito},
  {Dainotti}, {Guida}, \& {Ruffini}}]{2008AIPC..966....7B}
{Bernardini}, M.~G., {Bianco}, C.~L., {Caito}, L., {et~al.} 2008, in American
  Institute of Physics Conference Series, Vol. 966, Relativistic Astrophysics,
  ed. C.~L. {Bianco} \& S.~S. {Xue}, 7--11

\bibitem[{{Bernardini} {et~al.}(2005){Bernardini}, {Bianco}, {Chardonnet},
  {Fraschetti}, {Ruffini}, \& {Xue}}]{2005ApJ...634L..29B}
{Bernardini}, M.~G., {Bianco}, C.~L., {Chardonnet}, P., {et~al.} 2005, ApJ,
  634, L29

\bibitem[{{Bianco} {et~al.}(2008){Bianco}, {Bernardini}, {Caito}, {Dainotti},
  {Guida}, \& {Ruffini}}]{2008AIPC..966...12B}
{Bianco}, C.~L., {Bernardini}, M.~G., {Caito}, L., {et~al.} 2008, in American
  Institute of Physics Conference Series, Vol. 966, Relativistic Astrophysics,
  ed. C.~L. {Bianco} \& S.~S. {Xue}, 12--15

\bibitem[{{Bianco} \& {Ruffini}(2004)}]{2004ApJ...605L...1B}
{Bianco}, C.~L. \& {Ruffini}, R. 2004, ApJ, 605, L1

\bibitem[{{Bianco} \& {Ruffini}(2005)}]{2005ApJ...620L..23B}
{Bianco}, C.~L. \& {Ruffini}, R. 2005, ApJ, 620, L23

\bibitem[{{Bianco} {et~al.}(2001){Bianco}, {Ruffini}, \&
  {Xue}}]{2001A&A...368..377B}
{Bianco}, C.~L., {Ruffini}, R., \& {Xue}, S.-S. 2001, A\&A, 368, 377

\bibitem[{{Caito} {et~al.}(2008){Caito}, {Bernardini}, {Bianco}, {Dainotti},
  {Guida}, \& {Ruffini}}]{2008AIPC..966...16C}
{Caito}, L., {Bernardini}, M.~G., {Bianco}, C.~L., {et~al.} 2008, in American
  Institute of Physics Conference Series, Vol. 966, Relativistic Astrophysics,
  ed. C.~L. {Bianco} \& S.~S. {Xue}, 16--20

\bibitem[{{Dainotti} {et~al.}(2007){Dainotti}, {Bernardini}, {Bianco}, {Caito},
  {Guida}, \& {Ruffini}}]{2007A&A...471L..29D}
{Dainotti}, M.~G., {Bernardini}, M.~G., {Bianco}, C.~L., {et~al.} 2007, A\&A,
  471, L29

\bibitem[{{Firmani} {et~al.}(2006){Firmani}, {Ghisellini}, {Avila-Reese}, \&
  {Ghirlanda}}]{2006MNRAS.370..185F}
{Firmani}, C., {Ghisellini}, G., {Avila-Reese}, V., \& {Ghirlanda}, G. 2006,
  MNRAS, 370, 185

\bibitem[{{Gehrels} {et~al.}(2004){Gehrels}, {Chincarini}, {Giommi}, {Mason},
  {Nousek}, {Wells}, {White}, {Barthelmy}, {Burrows}, {Cominsky}, {Hurley},
  {Marshall}, {M{\'e}sz{\'a}ros}, {Roming}, {Angelini}, {Barbier}, {Belloni},
  {Campana}, {Caraveo}, {Chester}, {Citterio}, {Cline}, {Cropper}, {Cummings},
  {Dean}, {Feigelson}, {Fenimore}, {Frail}, {Fruchter}, {Garmire}, {Gendreau},
  {Ghisellini}, {Greiner}, {Hill}, {Hunsberger}, {Krimm}, {Kulkarni}, {Kumar},
  {Lebrun}, {Lloyd-Ronning}, {Markwardt}, {Mattson}, {Mushotzky}, {Norris},
  {Osborne}, {Paczynski}, {Palmer}, {Park}, {Parsons}, {Paul}, {Rees},
  {Reynolds}, {Rhoads}, {Sasseen}, {Schaefer}, {Short}, {Smale}, {Smith},
  {Stella}, {Tagliaferri}, {Takahashi}, {Tashiro}, {Townsley}, {Tueller},
  {Turner}, {Vietri}, {Voges}, {Ward}, {Willingale}, {Zerbi}, \&
  {Zhang}}]{2004ApJ...611.1005G}
{Gehrels}, N., {Chincarini}, G., {Giommi}, P., {et~al.} 2004, ApJ, 611, 1005

\bibitem[{{Ghirlanda} {et~al.}(2004){Ghirlanda}, {Ghisellini}, \&
  {Lazzati}}]{2004ApJ...616..331G}
{Ghirlanda}, G., {Ghisellini}, G., \& {Lazzati}, D. 2004, ApJ, 616, 331

\bibitem[{{Leibundgut}(2001)}]{2001ARA&A..39...67L}
{Leibundgut}, B. 2001, ARAA, 39, 67

\bibitem[{{Liang} \& {Zhang}(2005)}]{2005ApJ...633..611L}
{Liang}, E. \& {Zhang}, B. 2005, ApJ, 633, 611

\bibitem[{{Norris} \& {Bonnell}(2006)}]{2006ApJ...643..266N}
{Norris}, J.~P. \& {Bonnell}, J.~T. 2006, ApJ, 643, 266

\bibitem[{{Riess} {et~al.}(2001){Riess}, {Nugent}, {Gilliland}, {Schmidt},
  {Tonry}, {Dickinson}, {Thompson}, {Budav{\'a}ri}, {Casertano}, {Evans},
  {Filippenko}, {Livio}, {Sanders}, {Shapley}, {Spinrad}, {Steidel}, {Stern},
  {Surace}, \& {Veilleux}}]{2001ApJ...560...49R}
{Riess}, A.~G., {Nugent}, P.~E., {Gilliland}, R.~L., {et~al.} 2001, ApJ, 560,
  49

\bibitem[{{Ruffini} {et~al.}(2007){Ruffini}, {Bernardini}, {Bianco}, {Caito},
  {Chardonnet}, {Dainotti}, {Fraschetti}, {Guida}, {Rotondo}, {Vereshchagin},
  {Vitagliano}, \& {Xue}}]{2007AIPC..910...55R}
{Ruffini}, R., {Bernardini}, M.~G., {Bianco}, C.~L., {et~al.} 2007, in American
  Institute of Physics Conference Series, Vol. 910, XIIth Brazilian School of
  Cosmology and Gravitation, ed. M.~{Novello} \& S.~E. {Perez Bergliaffa},
  55--217

\bibitem[{{Ruffini} {et~al.}(2006){Ruffini}, {Bernardini}, {Bianco},
  {Chardonnet}, {Fraschetti}, {Guida}, \& {Xue}}]{2006ApJ...645L.109R}
{Ruffini}, R., {Bernardini}, M.~G., {Bianco}, C.~L., {et~al.} 2006, ApJ, 645,
  L109

\bibitem[{{Ruffini} {et~al.}(2004){Ruffini}, {Bianco}, {Chardonnet},
  {Fraschetti}, {Gurzadyan}, \& {Xue}}]{2004IJMPD..13..843R}
{Ruffini}, R., {Bianco}, C.~L., {Chardonnet}, P., {et~al.} 2004, IJMPD, 13, 843

\bibitem[{{Ruffini} {et~al.}(2005{\natexlab{a}}){Ruffini}, {Bianco},
  {Chardonnet}, {Fraschetti}, {Gurzadyan}, \& {Xue}}]{2005IJMPD..14...97R}
{Ruffini}, R., {Bianco}, C.~L., {Chardonnet}, P., {et~al.} 2005{\natexlab{a}},
  IJMPD, 14, 97

\bibitem[{{Ruffini} {et~al.}(2003){Ruffini}, {Bianco}, {Chardonnet},
  {Fraschetti}, {Vitagliano}, \& {Xue}}]{2003AIPC..668...16R}
{Ruffini}, R., {Bianco}, C.~L., {Chardonnet}, P., {et~al.} 2003, in American
  Institute of Physics Conference Series, Vol. 668, Cosmology and Gravitation,
  ed. M.~{Novello} \& S.~E. {Perez Bergliaffa}, 16--107

\bibitem[{{Ruffini} {et~al.}(2001){Ruffini}, {Bianco}, {Chardonnet},
  {Fraschetti}, \& {Xue}}]{2001ApJ...555L.113R}
{Ruffini}, R., {Bianco}, C.~L., {Chardonnet}, P., {Fraschetti}, F., \& {Xue},
  S.-S. 2001, ApJ, 555, L113

\bibitem[{{Ruffini} {et~al.}(2002){Ruffini}, {Bianco}, {Chardonnet},
  {Fraschetti}, \& {Xue}}]{2002ApJ...581L..19R}
{Ruffini}, R., {Bianco}, C.~L., {Chardonnet}, P., {Fraschetti}, F., \& {Xue},
  S.-S. 2002, ApJ, 581, L19

\bibitem[{{Ruffini} {et~al.}(2005{\natexlab{b}}){Ruffini}, {Fraschetti},
  {Vitagliano}, \& {Xue}}]{2005IJMPD..14..131R}
{Ruffini}, R., {Fraschetti}, F., {Vitagliano}, L., \& {Xue}, S.-S.
  2005{\natexlab{b}}, IJMPD, 14, 131

\bibitem[{{Ruffini} {et~al.}(2000){Ruffini}, {Salmonson}, {Wilson}, \&
  {Xue}}]{2000A&A...359..855R}
{Ruffini}, R., {Salmonson}, J.~D., {Wilson}, J.~R., \& {Xue}, S.-S. 2000, A\&A,
  359, 855

\bibitem[{{Sakamoto} {et~al.}(2006){Sakamoto}, {Barbier}, {Barthelmy},
  {Cummings}, {Fenimore}, {Gehrels}, {Hullinger}, {Krimm}, {Markwardt},
  {Palmer}, {Parsons}, {Sato}, \& {Tueller}}]{2006ApJ...636L..73S}
{Sakamoto}, T., {Barbier}, L., {Barthelmy}, S.~D., {et~al.} 2006, ApJ, 636, L73

\bibitem[{{Tagliaferri} {et~al.}(2005){Tagliaferri}, {Antonelli}, {Chincarini},
  {Fern{\'a}ndez-Soto}, {Malesani}, {Della Valle}, {D'Avanzo}, {Grazian},
  {Testa}, {Campana}, {Covino}, {Fiore}, {Stella}, {Castro-Tirado},
  {Gorosabel}, {Burrows}, {Capalbi}, {Cusumano}, {Conciatore}, {D'Elia},
  {Filliatre}, {Fugazza}, {Gehrels}, {Goldoni}, {Guetta}, {Guziy}, {Held},
  {Hurley}, {Israel}, {Jel{\'{\i}}nek}, {Lazzati}, {L{\'o}pez-Echarri},
  {Melandri}, {Mirabel}, {Moles}, {Moretti}, {Mason}, {Nousek}, {Osborne},
  {Pellizza}, {Perna}, {Piranomonte}, {Piro}, {de Ugarte Postigo}, \&
  {Romano}}]{2005A&A...443L...1T}
{Tagliaferri}, G., {Antonelli}, L.~A., {Chincarini}, G., {et~al.} 2005, A\&A,
  443, L1

\bibitem[{{Vaughan} {et~al.}(2006){Vaughan}, {Goad}, {Beardmore}, {O'Brien},
  {Osborne}, {Page}, {Barthelmy}, {Burrows}, {Campana}, {Cannizzo}, {Capalbi},
  {Chincarini}, {Cummings}, {Cusumano}, {Giommi}, {Godet}, {Hill}, {Kobayashi},
  {Kumar}, {La Parola}, {Levan}, {Mangano}, {M{\'e}sz{\'a}ros}, {Moretti},
  {Morris}, {Nousek}, {Pagani}, {Palmer}, {Racusin}, {Romano}, {Tagliaferri},
  {Zhang}, \& {Gehrels}}]{2006ApJ...638..920V}
{Vaughan}, S., {Goad}, M.~R., {Beardmore}, A.~P., {et~al.} 2006, ApJ, 638, 920

\bibitem[{{Yonetoku} {et~al.}(2004){Yonetoku}, {Murakami}, {Nakamura},
  {Yamazaki}, {Inoue}, \& {Ioka}}]{2004ApJ...609..935Y}
{Yonetoku}, D., {Murakami}, T., {Nakamura}, T., {et~al.} 2004, ApJ, 609, 935

\end{thebibliography}
\end{document}